# MEASUREMENTS OF ANISOTROPY IN THE COSMIC MICROWAVE BACKGROUND RADIATION AT 0.5 DEGREE ANGULAR SCALES NEAR THE STAR GAMMA URSAE MINORIS

M. J. Devlin[1], A. C. Clapp[1], J. O. Gundersen[2], C. A. Hagmann[1], V. V. Hristov[1],
A. E. Lange[1], M. A. Lim[2], P. M. Lubin[2], P. D. Mauskopf[1], P. R. Meinhold[2],
P. L. Richards[1], G. F. Smoot[3], S. T. Tanaka[1], P. T. Timbie[1,4],
and C. A. Wuensche[2,5]

[1] Physics Department, University of California at Berkeley, Berkeley, CA 94720; also NSF Center for Particle Astrophysics.

[2] Physics Department, University of California at Santa Barbara, Santa Barbara, CA 93106; also NSF Center for Particle Astrophysics.

[3] Physics Department, Lawrence Berkeley Laboratory, Berkeley, CA 94720; also NSF Center for Particle Astrophysics.

[4] Current address: Department of Physics, Box 1843, Brown University, Providence, RI 02912

[5] Also the Instituto Nacional de Pesquisas Espaciais - INPE/MCT, Departamento de Astrofísica, São José dos Campos, SP, Brazil 12200

## ABSTRACT

We present results from a four frequency observation of a 6° x 0°.6 strip of the sky centered near the star Gamma Ursae Minoris during the fourth flight of the Millimeter-wave Anisotropy eXperiment (MAX). The observation was made with a 1°.4 peak-to-peak sinusoidal chop in all bands. The FWHM beam sizes were $0°.55 \pm 0°.05$ at 3.5 cm$^{-1}$ and $0°.75 \pm 0°.05$ at 6, 9, and 14 cm$^{-1}$. During this observation significant correlated structure was observed at 3.5, 6 and 9 cm$^{-1}$ with amplitudes similar to those observed in the GUM region during the second and third flights of MAX. The frequency spectrum is consistent with CMB and inconsistent with thermal emission from interstellar dust. The extrapolated amplitudes of synchrotron and free-free emission are too small to account for the amplitude of the observed structure. If all of the structure is attributed to CMB anisotropy with a Gaussian autocorrelation function and a coherence angle of 25', then the most probable values of $\Delta T/T_{CMB}$ in the 3.5, 6, and 9 cm$^{-1}$ bands are $4.3^{+2.7}_{-1.6} \times 10^{-5}$, $2.8^{+4.3}_{-1.1} \times 10^{-5}$, and $3.5^{+3.0}_{-1.6} \times 10^{-5}$ (95% confidence upper and lower limits), respectively.

*Subject headings:* Cosmic microwave background - cosmology: observations

1. INTRODUCTION

Measurements of the anisotropy of the cosmic microwave background (CMB) provide an effective method for testing and constraining models of cosmic structure formation. The *Cosmic Background Explorer* (COBE) has detected anisotropy at large angular scales (>5°) (Smoot et al. 1992). Anisotropy measurements on medium angular scales constrain models of large-scale structure formation and the values of certain global parameters in cosmic evolution models (Bond et al. 1991; Vittorio et al. 1991). More powerful tests of cosmological models are possible when these measurements are combined with the measurements from COBE.

There has been a concerted recent effort to measure medium scale anisotropy (Fischer et al. 1992; Gaier et al. 1992; Schuster et al. 1993; Meinhold et al. 1993a; Wollack et al. 1993, Cheng et al. 1994, deBernardis et al. 1994). The results of CMB observations centered near the star Gamma Ursae Minoris (GUM) during the second and third flights of the Millimeter-wave Anisotropy eXperiment (MAX) were reported in Alsop et al. (1992) and Gundersen et al. (1993). We report here on a third CMB observation near GUM that occurred during the fourth flight of MAX.

2. INSTRUMENT

The instrument has been described in detail elsewhere (Fischer et al. 1992; Alsop et al. 1992; Meinhold et al. 1993b). It consists of an off-axis Gregorian telescope and a bolometric photometer mounted on an attitude-controlled balloon platform which attains altitudes of 36 km. The telescope has a one meter off-axis parabolic primary with a nutating elliptical secondary which modulates the beam sinusoidally in azimuth at 5.4 Hz with a peak-to-peak throw of 1°.4. A new single pixel, four band bolometric receiver was used for this flight. It features greatly reduced sensitivity to radio frequency (RF) interference, an additional 3.5 cm$^{-1}$ band, and an adiabatic demagnetization refrigerator to cool the bolometric photometer to 85 mK. The underfilled optics provide a $0°.55 \pm 0°.05$ FWHM beam in the single mode 3.5 cm$^{-1}$ band and $0°.75 \pm 0°.05$ FWHM beams in the multi-mode 6, 9, and 14 cm$^{-1}$ bands. The bandwidths are $\delta\nu/\nu$ = 0.57, 0.45, 0.35 and 0.25 FWHM, respectively. In order to convert antenna temperature measured in the 3.5, 6, 9, and 14 cm$^{-1}$ bands to a 2.73 K thermodynamic temperature, one must multiply the antenna temperatures by 1.54, 2.47, 6.18, and 34.1, respectively.

## 3. OBSERVATION

The instrument was launched from the National Scientific Balloon Facility in Palestine, TX at 2.5 UT 1993 June 15. CMB observations were made in three regions of the sky. The one reported here was made in a region near GUM $\alpha = 15^{\text{h}}20^{\text{m}}.7$, $\delta = 71°50'$ (Epoch 1993). The other two scans are reported in Clapp et al. (1994). The complete GUM scan lasted for 0.88 hours between UT = 5.97 hours and UT = 6.85 hours 1993 June 15. Calibrations were made before and after the observation using the membrane transfer standard described in Fischer et al. (1992). Scans of Jupiter were made between UT = 5.08 hours and UT = 5.33 hours to measure the beam size and to confirm the calibration. Using the best fit beam size, the derived temperature of Jupiter in each of the bands using the membrane calibration agrees with spectrum of Jupiter from Griffen et al (1986) to within 10%. Therefore we assume a 10% uncertainty in %%calibration. The instrument is calibrated so that a chopped beam centered between sky regions with temperatures $T_1$ and $T_2$ would yield $\Delta T = T_1 - T_2$. CMB observations consisted of constant velocity scans in azimuth of $\pm 3°.0$ on the sky while tracking GUM. A complete scan from 3° to -3° and back to 3° required 108 seconds.

## 4. DATA REDUCTION AND ANALYSIS

Transients due to cosmic rays were removed using an algorithm described in Alsop et al. (1992). This excluded 15 - 20% of the data. The detector %%output was demodulated using the sinusoidal reference from the nutating secondary to produce antenna temperature differences, $\Delta T_A$, on the sky. Data sets were produced that are in phase with the optical signal (optical phase) and 90° out of phase. The noise averaged over the entire GUM observation gives effective sensitivities of 631, 512, 792, and 2954 $\mu K\sqrt{sec}$ in CMB thermodynamic units in the 3.5, 6, 9, and 14 $cm^{-1}$ bands, respectively.

The signal in each band is significantly offset from zero. The average of the measured instrumental offsets in antenna temperature were 4.1, 1.1, 1.0, and 1.0 mK in the 3.5, 6, 9, and 14 $cm^{-1}$ bands. Tests done during the third flight of MAX where a temperature gradient was driven across the primary mirror indicate that the instrumental offset is primarily due to emission from the primary mirror. Variations were observed in the offsets in the 9 and 14 $cm^{-1}$ bands with a peak-to-peak amplitude of 0.2 to 0.3 mK on time scales of approximately 300 seconds. These offset variations are consistent with variations observed in previous flights and are attributed to atmospheric emission which increases rapidly with frequency. The offset in the 6 $cm^{-1}$ band varied with a peak-to-peak amplitude of 0.2 mK on time scales of approximately 150 seconds due to a malfunction of the 6 $cm^{-1}$ amplifier. No offset drift was observed in the 3.5 $cm^{-1}$ band. The offset and offset drifts were removed by performing a linear least squares fit to each half scan from 3° to -3° and from -3° to 3° in each of the bands. This removes any gradient in the astrophysical signal. The effect of this procedure was to greatly reduce the noise in the 6, 9, and 14 $cm^{-1}$ bands. The reduction in the rms amplitude of the signals in these

Preprint 4/15/94

bands was consistent with the reduction in the noise. The procedure had no effect on the noise or rms amplitude in the 3 cm$^{-1}$ band.

The means, variances and $1\sigma$ error bars of the antenna temperature differences were calculated for 63 pixels with approximately equal area separated by 17' in azimuth. The 63 pixels were formed by dividing the data set into three roughly equal time periods and producing a 21 pixel azimuthal scan for each period. This is done to minimize the effect of sky rotation. A map containing all pixel positions, means and error bars is available from the authors. All of the analysis described here will refer to the 63 pixel data set. Most of the essential features of the data are apparent in Figure 1 which shows the first two of the three time periods (which have significant spatial overlap) averaged into 21 pixels in azimuth. First, there is statistically significant structure in the 3.5, 6, 9 and 14 cm$^{-1}$ bands. Values of the rms $\Delta T_A$ obtained from the square root of the difference between the variance of the 63 pixel means and the variance due to detector noise are $81 \pm 16$, $28 \pm 6$, $10 \pm 4$, and $9 \pm 3$ $\mu K$ for the 3.5, 6, 9, and 14 cm$^{-1}$ bands, respectively. The error on each rms includes a $1\sigma$ statistical error and a $\pm 10\%$ estimate for the uncertainty in the absolute calibration. The probability of reproducing the measured rms with Gaussian random noise is negligibly small at 3.5 and 6 cm$^{-1}$, $3 \times 10^{-2}$ at 9 cm$^{-1}$, and $6 \times 10^{-4}$ at 14 cm$^{-1}$. By comparison, the probability for the 90° out of phase data is 0.86, $6 \times 10^{-2}$, 0.37, and 0.57 respectively. Second, the data at 3.5, 6, and 9 cm$^{-1}$ are significantly correlated with each other, but not with the data at 14 cm$^{-1}$. Third and last, the amplitude of the structure as measured in antenna temperature decreases with increasing frequency.

In order to test the hypothesis that the structure in the 3.5, 6, and 9 cm$^{-1}$ bands is correlated, a best fit model was determined by minimizing

$$\chi_R^2 = \sum_{j=1}^{3} \sum_{i=1}^{63} \left(x_{ij} - a_j y_i\right)^2 \bigg/ \sigma_{ij}^2. \qquad (1)$$

Here $x_{ij}$ and $\sigma_{ij}$ are the measured means and variances of the 63 pixels, $y_i$ represents the best-fit sky model, and $a_j$ are the best fit model scale factors for the 3.5, 6, and 9 cm$^{-1}$ bands. Three band fits were made using data for the 3.5, 6, and 9 cm$^{-1}$ bands. In addition, two band fits were made using data from various pairs of bands. The probability of exceeding the residual $\chi_R^2$ for each of the fits is given in column 2 of Table 1. In the case of the three band fit, the probability is 0.43 indicating that essentially all of the signal in excess of noise in the three bands is correlated and consistent with a single source of emission. The fits of the 3.5, 6 and 9 cm$^{-1}$ bands to the 14 cm$^{-1}$ band yield probabilities of <0.02 indicating that the signal in excess of noise in the 14 cm$^{-1}$ band is not correlated to the others. Analysis of the 14 cm$^{-1}$ signal shows that it is time variable and may to be due to emission from the atmosphere or possibly to sidelobe response. As discussed below, no significant time variability was observed at lower frequencies.

Table 1 also lists ratios of best fit scale factors for various fits and compares them to models of sky emission. This comparison is complicated by the 0°.55 FWHM size of the beam in the 3.5 cm$^{-1}$ band compared with the 0°.75 FWHM beam in the other bands. The observed spectrum will depend on both the spectrum of the emission and the angular distribution of the emission. Because the beam chop and scan is only in one dimension, a simple smoothing of the 3.5 cm$^{-1}$ band to broaden its response to 0°.75 FWHM is not valid. The observations can be compared with any model source spectra given an assumption about the angular distribution. Such comparisons are summarized in columns 4 - 8 of Table 1. Fourier analysis of the angular distribution of the signal in the 3.5 cm$^{-1}$ band indicates that much of the power of the signal is on angular scales smaller than the beam.

## 5. POTENTIAL SYSTEMATIC ERRORS

All anisotropy experiments are potentially susceptible to off-axis response from a variety of sources including the Sun, the Moon, the Earth, the balloon, and the Galactic plane. The unchopped off-axis response has been measured in the 3.5 cm$^{-1}$ band to be $\geq$ 65 dB below the on-axis response at angles from 13° to 35° in elevation below the boresight. No comparably deep measurements have been made of the chopped sidelobe response in azimuth. The Sun and Moon were both below the horizon. The elevation angle varied from 48°.5 to 45°.6 and the azimuthal position of the center of the scan varied from 347°.8 to 343°.3. Since the data set is divided into three roughly equal time periods, the elevation and azimuth of the center of the scan change by approximately 1° in each period. When the first two time periods, which overlap closely in sky coverage, are examined no significant change in the amplitude or angular distribution of the signal is seen. This implies that the signal remained fixed while the source of any sidelobe contamination has moved the equivalent of 7 pixels in Figure 1.

Baffles have been used in each flight of the MAX experiment to minimize the off-axis response of the instrument. The design of the baffles was modified for this flight to ensure that there was no direct or reflected emission from the Earth which could directly illuminate the optical system. The spectrum of the signal provides a strong constraint against sidelobe response. A sidelobe response resulting from reflections would result in a constant amplitude antenna temperature from the Earth and amplitude rising as $\nu^2$ from the balloon. In order to provide the observed spectrum, emission from the earth (or the balloon) which is diffracted into the beam, must undergo at least two (or four) diffractions.

This region of the sky has been observed three times each with substantially altered baffles and, in this observation, with different beam sizes. The elevation of GUM was about 30°, 49° and 44° in first, second and third observations, respectively. In each case a similar amplitude signal was observed with a spectrum consistent with CMB. It is unlikely that sidelobe contamination would be reproducible at this level under such drastically changed observing conditions.

## 6. GALACTIC AND EXTRAGALACTIC EMISSION

Potential Galactic and extragalactic sources of confusion include interstellar dust emission, synchrotron radiation and free-free emission. At 6, 9, and 14 cm$^{-1}$, the predominant Galactic and extragalactic source of confusion is expected to be emission from interstellar dust. The decrease in the amplitude of the observed structure as a function of frequency is not consistent with dust emission. In addition, based on an extrapolation from the IRAS 100 µm data (Wheelock et al. 1991) and scaling the brightness by the spectrum for high latitude dust reported by Meinhold et al. (1993a), the differential dust emission in the GUM region is expected to be a factor of 2 below the measured structure at 14 cm$^{-1}$ and a factor of 100 below the measured structure at 3.5 cm$^{-1}$.

The rms differential antenna temperature due to diffuse synchrotron emission in the GUM region is less than 0.8 K at 408 MHz as given in the 30' × 30' smoothed version of the Haslam et al. (1982) map. Assuming a scaling law $\Delta T_A \propto \nu^\beta$ for synchrotron emission with $\beta = -2.7$ implies an rms $\Delta T_A \leq 0.38, 0.06, 0.02,$ and $0.006$ µK at 3.5, 6, 9, and 14 cm$^{-1}$, respectively. These estimates account for <1% of the observed %%structure.

Free-free emission is the least well characterized of the potential Galactic contaminants. No small-scale Hα data are available for the GUM region. A catalogue search has been made for bright radio sources in this region. The beam intersects the source 3C314.1 which is expected to contribute less than 0.2 µK at 3.5 cm$^{-1}$ (Herbig & Readhead 1992). Estimates of diffuse free-free emission have been as described in Gundersen et al. (1993). Taking the conservative assumption that the entire 408 MHz rms is due to free-free emission and extrapolating to our frequencies using $\Delta T_A \propto \nu^{-2.1}$ gives

<15% of the measured rms. Preliminary measurements of a $2°.5 \times 2°.5$ region centered on GUM from the Cambridge Anisotropy Telescope (CAT) at 15 GHz using an interferometer with a $0°.5$ beam give no indication of a free-free component at 3.5 cm$^{-1}$ as large as the structure observed by MAX (Lasenby et al. 1993).

## 7. DISCUSSION

Since all of the known possible foreground contaminants are considered unlikely and the spectrum of the structure is consistent with CMB anisotropy, the following discussion interprets the observed structure as CMB anisotropy.

The rms of the data give $\Delta T_{rms}/T_{CMB} = 4.6 \pm 0.9 \times 10^{-5}$, $2.5 \pm 0.6 \times 10^{-5}$ and $2.2 \pm 0.8 \times 10^{-5}$ (68% confidence level) in the 3.5, 6, and 9 %%cm$^{-1}$ bands, respectively. This rms depends on the spatial window function determined by the beam size and chop. Comparison with theory requires integrating the predicted power spectrum of CMB anisotropy over the window functions. For a Cold Dark Matter (CDM) model with $\Omega_B = 0.03$ and h = 0.5 (Sugiyama & Gouda 1993) normalized to the rms amplitude of the anisotropy measured by COBE at large angular scales (Smoot et al. 1992), we expect $\Delta T_{rms}/T_{CMB} = 2 \times 10^{-5}$ for the 3.5 cm$^{-1}$ band and $\Delta T_{rms}/T_{CMB} = 1.6 \times 10^{-5}$ for the 6 and 9 cm$^{-1}$ bands. In comparing the predicted and measured values of $\Delta T_{rms}/T_{CMB}$, we should include an additional 20% uncertainty due to sampling %%variance (Scott et al. 1994).

If the CMB anisotropy is assumed to have a Gaussian autocorrelation function with a coherence angle of 25', then the most probable values of $\Delta T/T_{CMB}$ (with 95% confidence upper and lower limits) in the 3.5, 6, and 9 cm$^{-1}$ bands are $4.3^{+2.7}_{-1.6} \times 10^{-5}$, $2.8^{+4.3}_{-1.1} \times 10^{-5}$, and $3.5^{+3.0}_{-1.6} \times 10^{-5}$, respectively. The size of the observed structure is similar to that observed in the second and third flights of MAX where most probable

Preprint 4/15/94

amplitudes of $4.5^{+5.7}_{-2.6} \times 10^{-5}$ and $4.2^{+1.7}_{-1.1} \times 10^{-5}$ were obtained (Alsop et al. 1992, Gundersen et al. 1993, respectively). A direct comparison of the morphology of this data set with the results from the second and third flights is not possible since the scans only overlapped in a 0.5 deg$^2$ area.

## 8. CONCLUSION

We have presented new results from a search for CMB anisotropy with high sensitivity at angular scales near 1 degree. Significant structure is detected in the 3.5, 6 and 9 cm$^{-1}$ bands. Based on spectral and temporal arguments, sidelobe contamination from the Earth, balloon and the Galaxy are considered unlikely to cause the observed structure. The data rule out Galactic dust emission via the spectrum, morphology and amplitude of the structure. Synchrotron and free-free emission are considered unlikely contaminants from estimates of their intensity based on low-frequency maps. The relative amplitude of the structure at 3.5, 6, 9 and 14 cm$^{-1}$ is consistent with the CMB spectrum. This is the third time that the MAX experiment has measured structure with CMB spectrum over a broad spectral range with similar amplitude in the GUM region (Alsop et al. 1992, Gundersen et al. 1993). While the experiment has been significantly changed the result has remained the same. It has become increasingly difficult to construct alternative hypotheses for the observed signals.

This work was supported by the National Science Foundation through the Center for Particle Astrophysics (cooperative agreement AST-9120005), the National Aeronautics and Space Administration under grants NAGW-1062 and FD-NAGW-2121, the University of California, and previously California Space Institute. The work of C. A. Wuensche was partially supported by a fellowship from the Conselho Nacional de Desenvolvimento Científico e Tecnológico (CNPq), Brazil, under grant 200833/91.0. The authors would like to thank Geoff Cook for hardware support and Douglas Scott and Martin White for useful discussions.

TABLE 1

Ratios of the Best Fit Model Scale Factors for Pairs of Bands Compared

with Ratios Computed from Various Theoretical Models of Emission

| Band Ratio | Probability | Scale Factor Ratios [c] | CMB [d] | CMB CDM[e] | CMB Unresolved Source | Free-Free[d] | Free-Free Unresolved Source |
|---|---|---|---|---|---|---|---|
| (1) | (2) | (3) | (4) | (5) | (6) | (7) | (8) |
| 6/3.5[a] | } 0.43 { | 0.37 ±0.09 | 0.62 | 0.50 | 0.34[f] | 0.38 | 0.20[f] |
| 9/3.5[a] |  | 0.17 ±0.04 | 0.25 | 0.20 | 0.13[f] | 0.16 | 0.08[f] |
| 6/3.5[b] | 0.13 | 0.39 ±0.10 | 0.62 | 0.50 | 0.34[f] | 0.38 | 0.20[f] |
| 9/3.5[b] | 0.90 | 0.17 ±0.05 | 0.25 | 0.20 | 0.13[f] | 0.16 | 0.08[f] |
| 9/6[b] | 0.51 | 0.40 ±0.14 | 0.40 | 0.40 | 0.40 | 0.42 | 0.42 |

[a] Three band fit. [b] Two band fit. [c] Ratio of antenna temperatures. Best-fit ratio ± 1 $\sigma$ error. This includes a 10% calibration error. [d] Ratios calculated with no beam size correction. [e] Ratios calculated using Cold Dark Matter model with $\Omega_B = 0.03$ and h = 0.5. [f] Point source response. This is considered a lower limit.

REFERENCES

Alsop, D. C., et al. 1992, ApJ, 395, 317

Bond, J. R., et al. 1991, Phys. Rev. Lett., 66, 2179

Cheng, E. S., et al. 1994, ApJ, 422, L37

Clapp, A. C., et al. 1994, submitted ApJ

deBernardis, P., et al. 1994, ApJ, in press

Fischer, M., et al. 1992, ApJ, 388, 242

Gaier, T., et al. 1992, 398, L1

Griffen, M. J., et al. 1986, Icarus, 65, 244

Gundersen, J. O., et al. 1993, ApJ, 413, L1

Haslam, C. G. T., et al. 1982, A&AS, 47, 1

Herbig, T., Readhead, A. C. S. 1992, ApJ Supl. Ser., 81, 83

Lasenby, A., et al. 1993, personal communication

Meinhold, P. R., et al. 1993a, ApJ, 409, L1

Meinhold, P. R., et al. 1993b, ApJ, 406, 12

Schuster, J., et al. 1993, ApJ, 412, L47

Scott, D., Srednicki, M., White, M. 1994, ApJ, 421, L5

Smoot, G. F., et al. 1992, ApJ, 396, L1

Sugiyama, N., Gouda, N. 1992, Prog. Theor. Phys., 88, 803

Vittorio, N., et al. 1991, ApJ, 372, L1

Wheelock, S., et al. 1991, IRAS Sky Survey Atlas

Wollack, E., et al. 1993, ApJ, 419, L49

# FIGURE CAPTIONS

FIG. 1.--- Antenna temperature differences (± 1 σ) for 30 minutes of data near GUM. Each point is separated by 17' in azimuth.

Postal Address Page

A. C. Clapp, M. J. Devlin, C. A. Hagmann, V. V. Hristov, A. E. Lange, P. D. Mauskopf
P. L. Richards, and S. T. Tanaka:

>Physics Department
>
>University of California at Berkeley
>
>Berkeley, CA 94720

J. O. Gundersen, M. A. Lim, P. M. Lubin, P. R. Meinhold, C. A. Wuensche

>Physics Department
>
>University of California at Santa Barbara
>
>Santa Barbara, CA 93106

G. F. Smoot

>Physics Department
>
>Lawrence Berkeley Laboratory
>
>Berkeley, CA 94720

P. T. Timbie

>Department of Physics
>
>Box 1843
>
>Brown University
>
>Providence, RI 02912

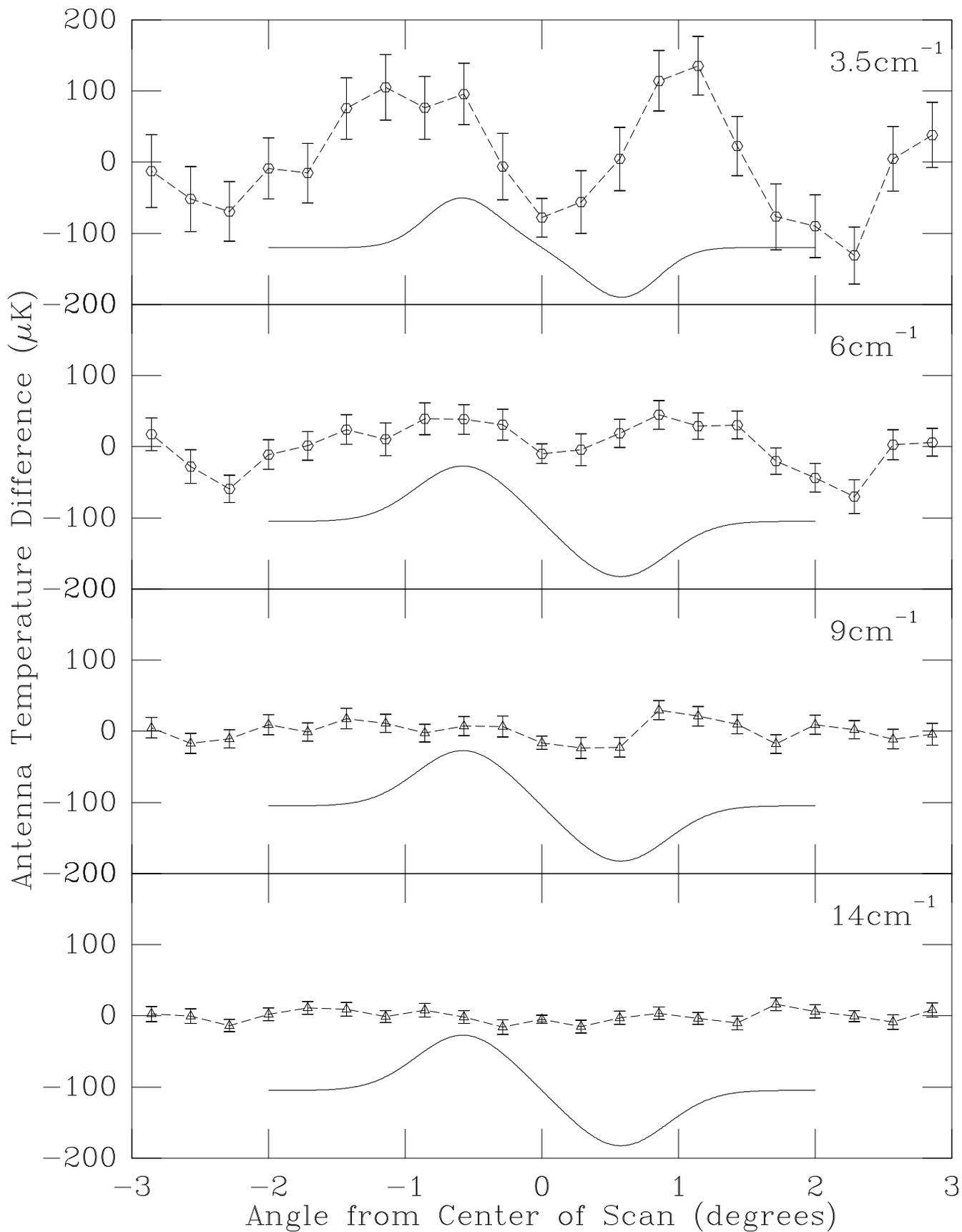